\documentclass{article}
\usepackage{hyperref}
\usepackage{xurl}
\usepackage{datetime}
\usepackage{graphicx}
\usepackage{natbib}
\usepackage{booktabs}
\hypersetup{colorlinks=true, allcolors=blue}

\title{Unveiling the Predators: Contemporary Approaches to Identifying Illegitimate Open Access Journals in the Academic Publishing Ecosystem}

\author{Robert Šamárek \and Radek Martinek}

\newdate{articleDate}{1}{9}{2025}
\date{\displaydate{articleDate}}

\begin{document}
\maketitle
\begin{abstract}Predatory journals pose a significant challenge to the integrity of the Open Access (OA) publishing model by exploiting its framework for financial gain while bypassing essential editorial and peer-review standards. This study critically evaluates existing methodologies for identifying such journals, ranging from manual blacklist checks to advanced automated approaches utilizing machine learning. The analysis highlights critical limitations, including the lack of a universally accepted definition of predatory journals, over-reliance on binary classification systems (e.g., blacklists and whitelists), and issues with scalability, reliability and interpretability. To address these shortcomings, this paper introduces a novel methodology based on multivariate graph analysis. By modeling the academic publishing ecosystem as a network of interconnected entities (such as authors, articles, journals, and publishers), this approach provides broader insights into the dynamics of scholarly communication and could help identify illegitimate publishing practices by utilizing graph algorithms like centrality measures, community detection, and anomaly detection. The proposed framework aims to enhance the accuracy, scalability, and transparency of detection of illegitimate journals and publishing practices while fostering a more comprehensive understanding of the academic publishing landscape.\end{abstract}

\section{Introduction}

\begin{figure}[!htbp]
\centering
\includegraphics[width=0.75\linewidth]{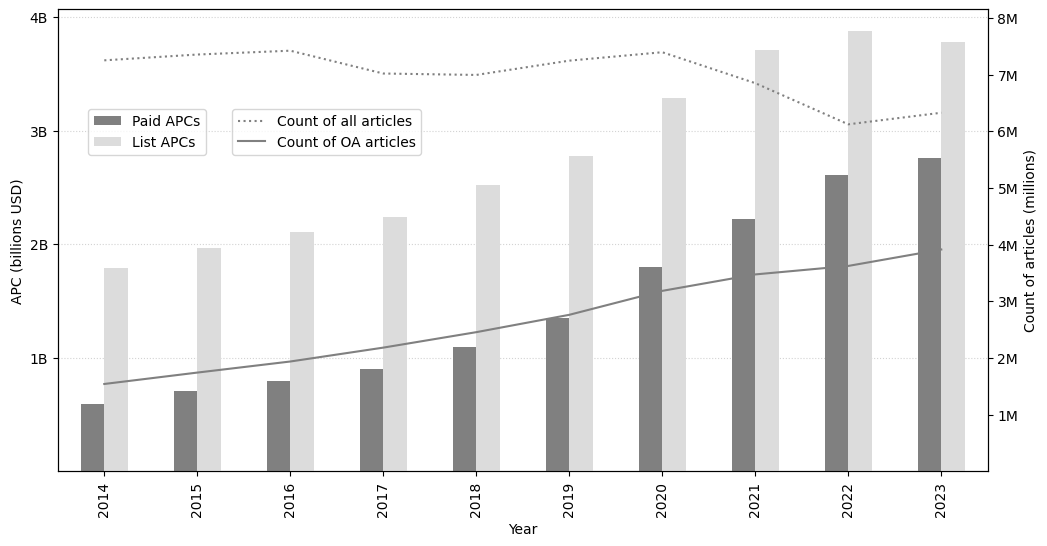}
\caption[]{Cumulative APCs paid by authors from 2014 to 2023

This chart illustrates the cumulative Article Processing Charges (APCs) paid by authors (and listed by journals) in USD from 2014 to 2023. The data show an exponential growth in the total amount of APCs paid over the years in comparison to linear growth in the number of OA articles published (from 1.5M in 2014 to 3.9M in 2023). This trend indicates a significant increase in the financial burden on authors and/or their institutions to publish research in OA journals.}
\label{apc_and_counts}
\end{figure}

Open Access (OA) publishing\citep{OpenAccess101}, strengthened by significant academic and government support in the early 2000s, has revolutionized the dissemination of scholarly work, allowing unrestricted access to research papers online. This model ensures that anyone can read and use these publications without the barrier of subscription fees, thus fostering wider knowledge sharing and collaboration across the globe. However, this business model relies on alternative funding strategies, often involving authors paying Article Processing Charges (APC) to cover the costs of publication. According to estimates based on OpenAlex\citep{priemOpenAlexFullyopenIndex2022} data\citep{openalex}, the annual cumulative APCs paid by authors in 2023 reached 2.76 billion USD which is a significant increase from 0.59 billion USD in 2014 (see Figure~\ref{apc_and_counts}). In this 10-year period, the publishing industry received 14.8 billion USD in APCs.

\begin{figure}[!htbp]
\centering
\includegraphics[width=0.75\linewidth]{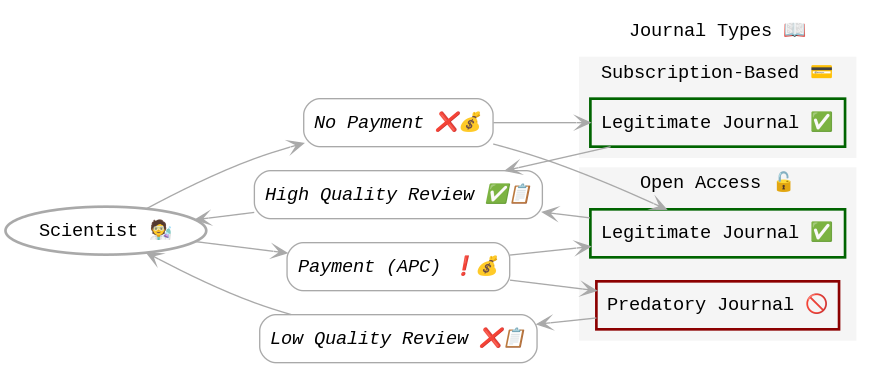}
\caption[]{Relationship Scientist - Journal

This diagram illustrates the relationship between a scientist and a journal, highlighting the publishing options available to the scientist. The scientist can choose between subscription-based journals, open access journals with no payment required, and open access journals with Article Processing Charges (APCs) to cover publication costs. It also underscores the risks associated with predatory journals that exploit the APC model without providing legitimate editorial services.}
\label{relationship-scientist-journal-reader-1-3}
\end{figure}

\begin{figure}[!htbp]
\centering
\includegraphics[width=0.85\linewidth]{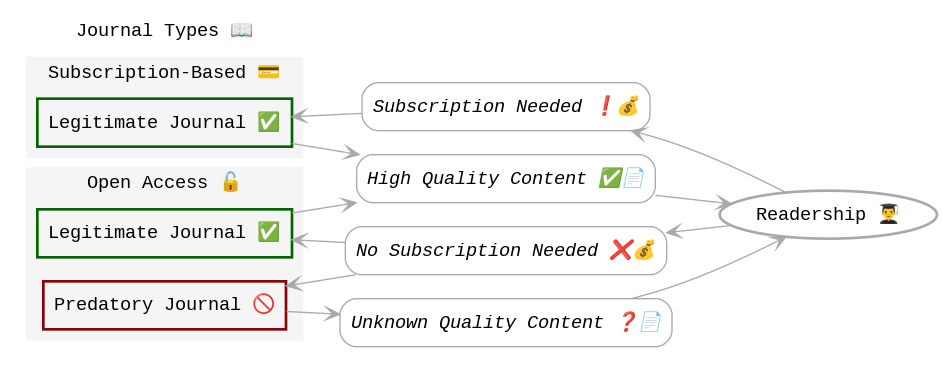}
\caption[]{Relationship Reader - Journal

This diagram shows the relationship between readership and journal, emphasizing access options and content quality. Legitimate journals, both subscription-based and open access, offer high-quality content, while predatory journals provide questionable content without subscription barriers.}
\label{relationship-scientist-journal-reader-3-5}
\end{figure}

While the OA concept has democratized the availability of research, it has also given rise to a new breed of publishers and journals that exploit the system for financial gain, without providing the necessary editorial and peer review services. These journals are commonly referred to as ``predatory'' and have become a significant concern for the academic community. Simplified diagrams illustrate the publishing options for a scientist (Figure~\ref{relationship-scientist-journal-reader-1-3}) and the consuming options for a reader (Figure~\ref{relationship-scientist-journal-reader-3-5}).

The terms ``predatory journal'' and ``predatory publishing'' were coined by Jeffrey Beall, a researcher and librarian at the University of Colorado, Denver. He started to maintain a list of predatory journals on his blog in 2008 and it became widely known (as the Beall's List\citep{BeallsList}) in the scientific community in 2010. The original list was maintained by Beall until January 2017, when the Scholarly Open Access website was taken offline; archived and independently maintained successor versions have remained available since then, and the term is still widely used.

This review aims to map out methodologies for identifying predatory journals, evaluate their effectiveness, and propose a new methodology to address existing gaps.

\section{Literature}

This section focuses on recent scholarly articles published within the last six years, particularly those that introduce new methodologies or critically assess existing ones. The goal is to provide a comprehensive understanding of the current approaches used to identify predatory journals.

\subsection{Terminology and Definitions}

Grudniewicz et al.\citep{grudniewiczPredatoryJournalsNo2019} state that ``\textit{barrier to combating predatory publishing is ... lack of an agreed definition}'', and propose the following:

\begin{quote}
Predatory journals and publishers are entities that prioritize self-interest at the expense of scholarship and are characterized by false or misleading information, deviation from best editorial and publication practices, a lack of transparency, and/or the use of aggressive and indiscriminate solicitation practices.
\end{quote}

This definition was reached by consensus of 43 experts from 10 countries during a workshop in Ottawa, Canada, in 2019.

Another definition is provided by Diane Pecorari\citep{pecorariPredatoryConferencesWhat2021} citing The Committee on Publication Ethics:

\begin{quote}
Predatory publishing generally refers to the systematic for-profit publication of purportedly scholarly content . . . in a deceptive or fraudulent way and without any regard for quality assurance. Here, `for-profit' refers to profit generation per se. Whereas predatory publishers are profit-generating businesses, some may conceivably pose as non-profit entities such as academic societies or research institutions. This is not to suggest that `for profit' is, in itself, problematic but that these journals exist solely for profit without any commitment to publication ethics or integrity of any kind.
\end{quote}

Ateeq et al.\citep{ateeqIntelligentFrameworkDetecting2023} also note the lack of agreement about the characteristics of predatory venues:

\begin{quote}
However, there is general agreement among some scholars to define predatory venues as venues that claim to conduct proper peer review, while in fact, they do not. Different scholars have tied different characteristics to this notion, such as spamming researchers by sending emails to publish, pretending to have sufficient quality control, poor editorial services and poor copyediting, and charging researchers with excessive and non-disclosed publication fees.
\end{quote}

For the purpose of this paper, we can take the minimal common denominator from the definitions above and consider predatory journals as: ``entities that prioritize self-interest at the expense of scholarship and are characterized by false or misleading information, deviation from best editorial and publication practices, a lack of transparency, and/or the use of aggressive and indiscriminate solicitation practices''.

Besides predatory journals, we should also mention predatory conferences, which use similar practices to predatory journals, and are also a significant issue in the academic community\citep{pecorariPredatoryConferencesWhat2021}.

Another term used in the context of unethical or fraudulent publishing practices is ``Hijacked Journals''. Unlike predatory journals, hijacked journals are legitimate journals whose websites have been mimicked (including the journal's name, logo, visual style, ISSN, etc.) by scammers to deceive authors into submitting their work and paying high fees. Another way to hijack a journal is to take over the journal's original URL domain\citep{abalkinaPredatoryVsHijacked2023}. These criminals then issue fake calls for paper submissions through emails and other channels, which are often difficult to distinguish from legitimate calls. There have been attempts to classify journal sites as hijacked or not using scraped data from the journal's website and a decision-tree algorithm\citep{andoohginshahriDetectingHijackedJournals2017}.

\subsection{Methodological Approaches to Identifying Predatory Journals}

Methodologies described in the literature can be classified based on the nature of data or information sources, as well as the analytical approaches (see Table~\ref{comparative-matrix}). The first category of sources is unstructured data about journals' review processes, editorial boards, and journal descriptions, usually harvested from journal websites manually or by web scraping. The second category is structured data, typically journal metadata (bibliometrics). Methods of analysis are categorized as manual and automated. Automated methods could be further divided into simple algorithms and machine learning algorithms.

\begin{table}
\centering
\caption[]{Comparative Matrix of Analytical Methods and Data Sources Utilized in Journal Evaluation

This table illustrates the dichotomy between manual and automated analysis methods across different types of data sources, highlighting the trade-offs between depth of insight, scalability, and interpretability.}
\label{comparative-matrix}
\begin{tabular}{p{\dimexpr 0.333\linewidth-2\tabcolsep}p{\dimexpr 0.333\linewidth-2\tabcolsep}p{\dimexpr 0.333\linewidth-2\tabcolsep}}
\toprule
Method {\textbackslash}~Data source & A. Publication process \& Journal description & B. Journal metadata (Bibliometrics) \\
\hline
1. Manual analysis & Involves detailed examination of journal processes and governance, requiring extensive human effort. Time-consuming and not scalable. & Deterministic approaches, such as citation analysis and impact factor evaluation, with direct interpretation of metrics. \\
2. Automated analysis & Utilizes Machine Learning (ML) and/or Large Language Models for deep content analysis, offering scalability but facing challenges in transparency and interpretability. & Employs statistical methods and/or ML algorithms to analyze bibliometric data, providing quantitative insights. \\
\bottomrule
\end{tabular}
\end{table}

Manual analysis based on investigation of a journal's processes and description (quadrant 1A in the Table~\ref{comparative-matrix}) is a common approach to identifying predatory journals. This method involves examining the journal's website, reviewing its editorial board, peer review process, publication fees, and adherence to ethical guidelines. The analysis is typically conducted by researchers who evaluate the journal's legitimacy based on a set of predefined criteria or by expert personnel (e.g., librarians). This approach is time-consuming and requires significant human effort, but it offers detailed insights into the journal's practices and can help identify potential red flags.

As a helper tool for manual analysis, researchers can use guidelines, checklists, and decision trees to evaluate the legitimacy of journals. An example of a decision tree is in \citep{richtigProblemsChallengesPredatory2018} and its simplified version in Appendix~1.

As a subcategory of manual analysis, we may consider simply checking the journal title against a whitelist or blacklist, e.g., DOAJ\citep{DOAJ} or Beall's List\citep{BeallsList}. While this approach is simple and quick, it is inherently limited in scope, as no single list can encompass the entirety of existing journals.

Because of the limitations of manual solutions, automated detections are being developed (2A). Ateeq et al.\citep{ateeqIntelligentFrameworkDetecting2023} reject manual analysis and propose an Intelligent Framework for Detecting Predatory Publishing Venues that utilizes artificial intelligence (AI) techniques. The authors created a dataset of 9,866 journals annotated as predatory or legitimate and then trained seven machine learning and deep learning models. The best results were achieved by a Convolutional Neural Network (CNN), with an F1 score of 0.96. The dataset for modeling was created as follows: data acquisition from DOAJ\citep{DOAJ}, Beall's List\citep{BeallsList}, and venues' websites (web scraped), legitimacy criteria compilation (39 criteria), and dataset annotation as legitimate or non-legitimate (those classified further by criteria violation).

A similar approach was taken by Chen et al. with their Academic Journal Predatory Detection (AJPD) system\citep{chenOpenAutomationSystemForPredatoryJournalDetection2023}. The authors picked journals' references from updated Beall's List and the Stop Predatory Journals list, identifying 833 predatory journal links, and legitimate journal data from the Directory of Open Access Journals (DOAJ) and the Berlin Institute of Health (BIH) Quest website, collecting 1213 legitimate journal links. The dataset for training was created by web scraping the journals' websites, removing HTML tags, and normalizing the text. Comparing eight classification algorithms, the Random Forest model exhibited the best performance in terms of both prediction (with a recall rate of 0.982 and an F1 score of 0.98).

But, as Teixeira da Silva et al. showed, the automated AI methods could be misleading. In the study ``Can AI Detect Predatory Journals? The Case of FT50 Journals''\citep{teixeiradasilvaCanAIDetect2023}, the authors utilized the AI tool (AJPD) from the paper by Chen et al.\citep{chenOpenAutomationSystemForPredatoryJournalDetection2023} to evaluate 50 journals from the prestigious Financial Times (FT50) list. This testing revealed that 88\% of these journals were classified as ``suspected predatory journals'' (SPJs), with only six being marked as ``normal journals'' (NJs).

A different approach was taken by Richtig et al. in their study\citep{richtigPredatoryJournalsPerception2023}. The authors utilized bibliometric data from sources such as the ISSN database, PubMed, Scopus, Web of Science, Crossref, and the Directory of Open Access Journals (DOAJ). They aimed not to classify journals as predatory but to assess the impact of Beall's list on authors' publication behavior. By analyzing bibliometric data from various scholarly databases, the study examined how being listed on Beall's list influenced researchers' perceptions and publishing choices.

Teixeira da Silva et al. suggest a Credit-like Rating methodology (inspired by the financial sector) to determine journals' legitimacy\citep{teixeiradasilvaCreditlikeRatingSystem2021}. This is another example of manual analysis of a journal's description and processes (1A). Rather than a binary classification, the authors propose ten levels of journals' risk-rating, from AAA (lowest risk) to D (extreme risk).

Referring to the B column of our table, studies utilizing bibliometric metadata analysis exclusively for detecting predatory journals remain scarce. Wu et al. \citep{wuIdentificationCausalAnalysis2024} addressed this gap by constructing an evaluation system based on journal metadata and bibliometric indicators. They developed an interpretable machine learning model that incorporates metrics such as the Impact Factor, Cited Half-Life, SNIP, and Self-Citation Rate, combined with SHAP analysis to identify journals with quality risks. While this approach leverages structured data effectively, its focus extends beyond detection, aiming to provide causal explanations for journal risk profiles rather than exclusively flagging predatory journals. However, the study highlights the need for model calibration to account for specific disciplinary categories, as the relevance and weight of bibliometric indicators vary significantly across different academic fields, influencing the model's effectiveness and reliability.

We should also mention the practices of citation databases. They usually do not address journals as predatory or not, but they do delist journals from their database indexes occasionally. Each database has its own process and criteria for delisting. For example, Clarivate's Web of Science refers to \textit{quality criteria}\citep{WoSCriteria} as mentioned in the article Supporting integrity of the scholarly record: Our commitment to curation and selectivity in the Web of Science\citep{WoSDelisting}. In 2023 Web of Science delisted 86 journals from their Core Collection index with justifications:

\begin{quote}
Journal has been re-evaluated and does not meet one or more of the quality criteria, resulting in removal from Web of Science Core Collection.
\end{quote}

Clarivate's \textit{Monthly Change Archive} is available (after login) for download as a spreadsheet file here\citep{WoSCollectionListDownloads}. Those files contain the list of journals that were added or removed from the Core Collection index in a given month.

\section{Challenges in Defining and Identifying Predatory Journals}

The academic community has been grappling with the phenomenon of predatory journals for over a decade, yet a consensus on defining and identifying these entities remains elusive.

\subsection{Implications of Definitions}

The lack of consensus on the definition of predatory journals and the absence of tools or criteria for their identification have led to confusion and inconsistency in identifying illegitimate journals.

It could be a reason why authors are still submitting their work to questionable
journals and citing them in their research. A study by Richtig et al.\citep{richtigPredatoryJournalsPerception2023} suggests that Beall's list's effectiveness in deterring submissions to these journals may be overestimated, indicating that authors may prioritize other criteria and databases when selecting publication venues.

\subsection{Critique of Binary Classification and the ``Predatory'' Label}

The binary classification of journals into ``predatory'' versus ``legitimate'' categories oversimplifies the complex nature of academic publishing and raises important ethical implications. The boundary between these two categories is not sharp but rather fuzzy, and the criteria for classification are not universally agreed upon. This approach not only risks misclassification but also may unjustly stigmatize journals on the fringes of these definitions. Moreover, the ``predatory'' label can inflict reputational damage not only on journals but also on authors and reviewers. Instead of labeling journals as ``predatory'', Grudniewicz et al. propose to use the terms ``illegitimate'', ``deceptive'', or ``acting in bad faith''\citep{grudniewiczPredatoryJournalsNo2019}.

There are alternatives to binary classification, such as multi-class classification\citep{teixeiradasilvaCreditlikeRatingSystem2021}, probabilistic classifiers, regression models or machine learning clustering, which may offer a more nuanced and accurate assessment of journal legitimacy.

\section{Limitations of Current Methodologies}

Current methodologies for identifying predatory journals, whether manual or automated, exhibit significant limitations that hinder their effectiveness and reliability.

Manual analysis, while offering detailed insights, is time-consuming and not scalable and suffers from lack of agreed-upon definitions and workflow. On the other hand, the simplicity and transparency of manual analysis are the main advantages of this approach. Manual analysis is also more flexible, does not rely on advanced technical skills and technology, and can be adapted to the specific needs of the researcher.

\begin{figure}[!htbp]
\centering
\includegraphics[width=0.75\linewidth]{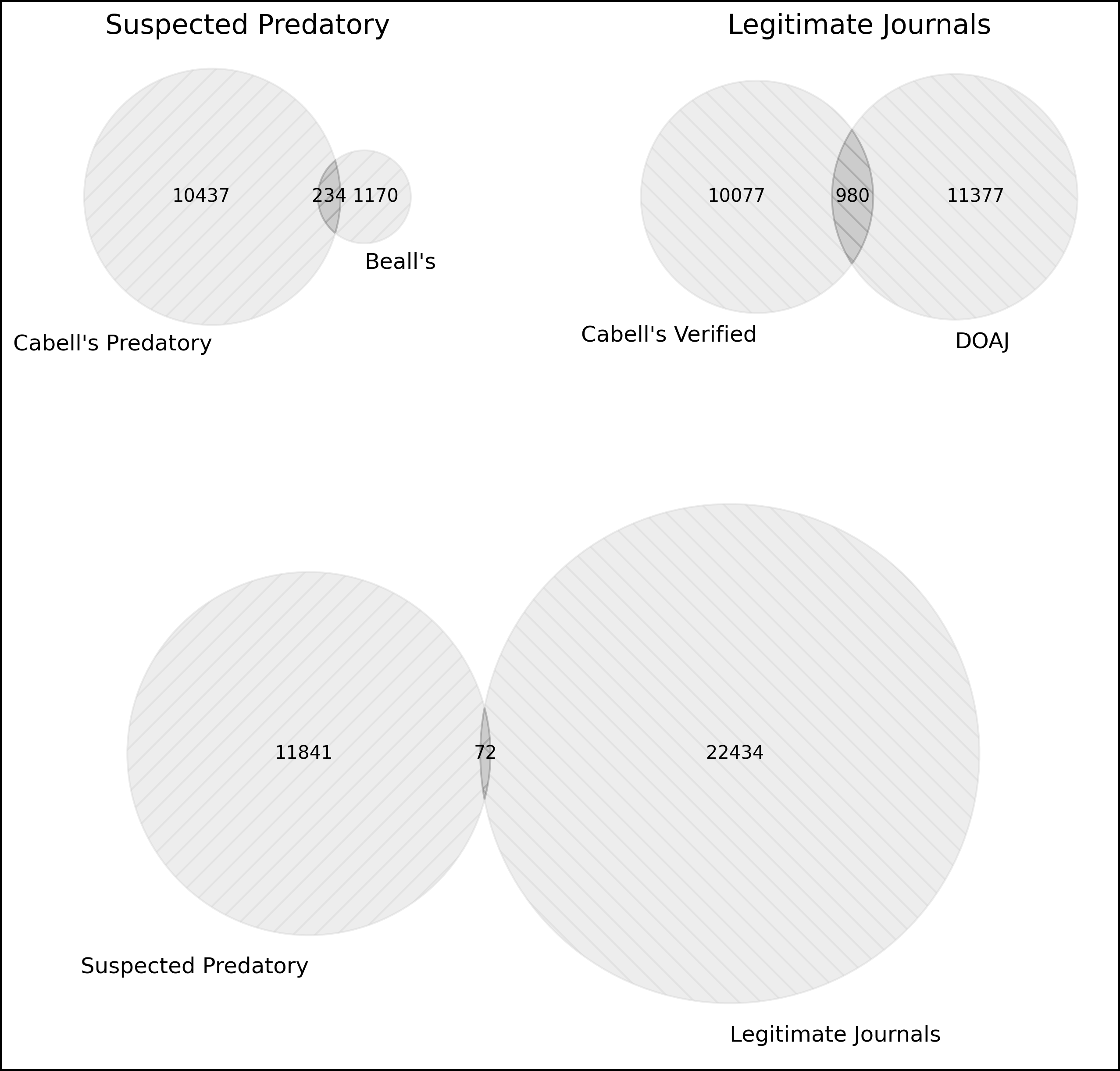}
\caption[]{The Overlap and Discrepancies Between Predatory and Legitimate Journal Listings

These Venn diagrams examine two blacklists (Beall's List and Cabells' blacklist) and two whitelists (DOAJ and Cabells' whitelist) to explore the overlap and discrepancies among them highlighting the inherent challenges in creating a comprehensive and non-contradictory classification system for open access journals. It reveals that some journals (72) are simultaneously deemed legitimate by one source and predatory by another.

This visualization underscores the limitations of relying solely on whitelist or blacklist checks to assess journal credibility, as these lists can both miss certain journals and include contradictory classifications.

Inspired by the Figure ``No List to Rule Them All''\citep{grudniewiczPredatoryJournalsNo2019}.}
\label{no-list-to-rule-them-all}
\end{figure}

Important limitations of blacklists and whitelists are their incompleteness, subjectivity, and lack of transparency in the criteria used for classification. Moreover, as highlighted by Grudniewicz et al.\citep{grudniewiczPredatoryJournalsNo2019}, these lists may sometimes present conflicting information (Figure~\ref{no-list-to-rule-them-all}).

Automated analyses, particularly those utilizing AI and machine learning, offer scalability and efficiency but face challenges related to accuracy, transparency and interpretability. Wu et al. \citep{wuIdentificationCausalAnalysis2024} employed the SHAP framework, which enhances model transparency and provides detailed explanations of risk factors. However, their study also underscores the challenge of calibrating models for specific disciplinary contexts. Another AI-based tool, despite claiming high precision, misclassified reputable journals from the FT50 list as predatory\citep{teixeiradasilvaCanAIDetect2023}, raising questions about AI reliability in this context. Furthermore, most existing AI tools are trained on mentioned blacklists and whitelists, thereby inheriting their limitations and biases.

A more fundamental limitation, increasingly apparent in the most recent evidence, is that all of the above methods operate primarily at the level of the \textit{individual} journal or article. Yet contemporary illegitimate publishing is organized as a \textit{network}: recent large-scale analysis shows that the entities enabling scientific fraud --- paper mills brokering authorship, coordinating submissions, and targeting indexed venues --- are large, resilient, and growing faster than corrective measures, spanning thousands of articles, hundreds of journals, and thousands of authors across many countries \citep{richardsonEntitiesEnablingScientific2025}. Generative AI further lowers the cost of producing superficially plausible manuscripts and fabricated reviewer personas at scale. Coordination of this kind is, by construction, a relational phenomenon: it leaves structural traces (shared authors, circular citation, clustered venues) that are invisible to indicators computed one journal at a time, but legible in the topology of the publishing network. This motivates a shift from per-entity classification toward analysis of the network itself.

\section{Discussion}

The limitations of existing methodologies underscore the need for a new approach to identifying illegitimate journals.

\subsection{From Lists to Relational Evidence: A Graph-Based Direction}

\begin{figure}[!htbp]
\centering
\includegraphics[width=0.75\linewidth]{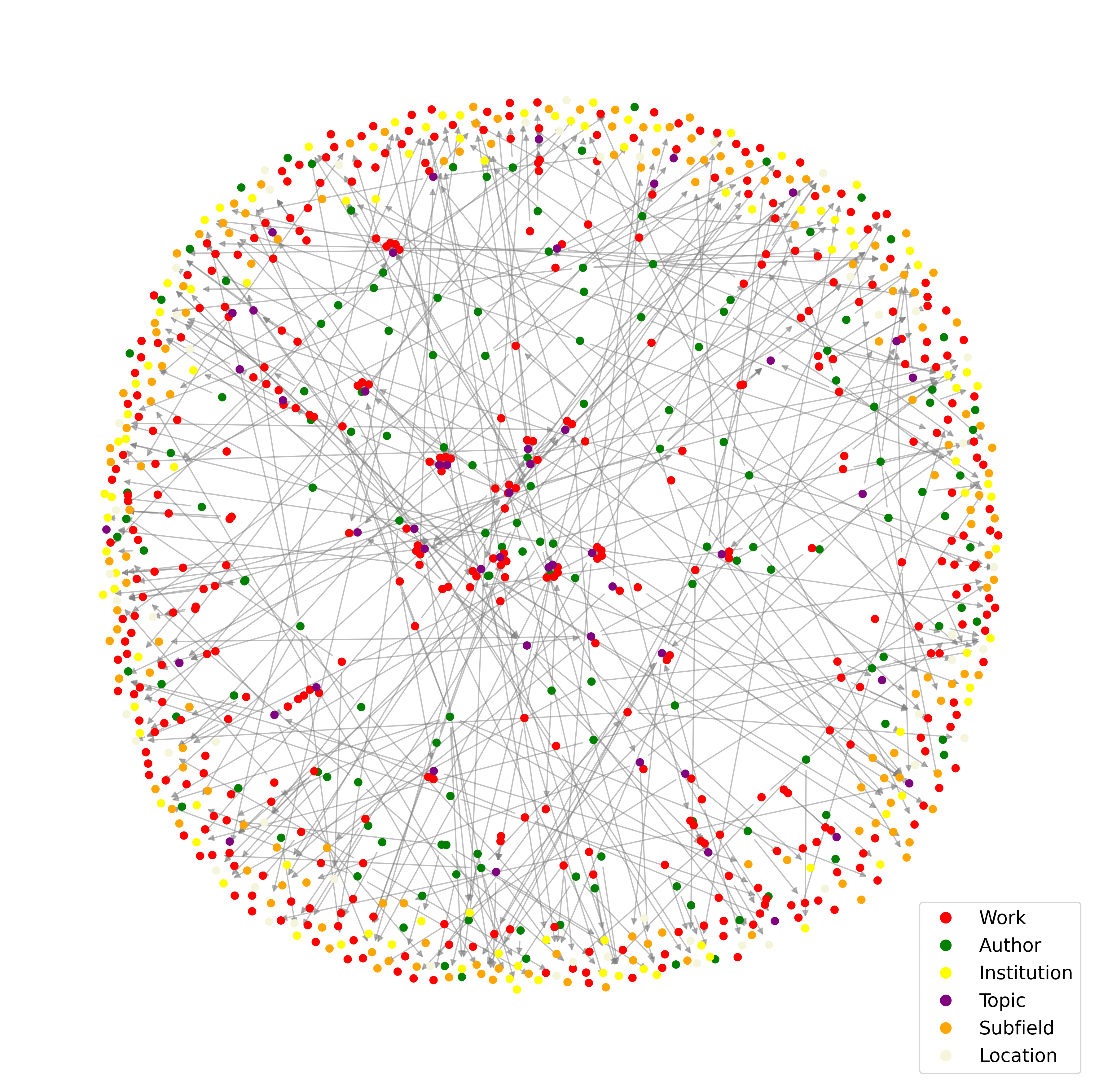}
\caption[]{Big Picture Graph Representation

This figure provides a conceptual overview of the proposed multivariate graph model, illustrating how different entities in the academic ecosystem are connected and creating patterns and communities. Each node represents an entity like author, journal, or work (paper) represented by a different color, and each edge represents a relationship (authorship, affiliation, topic association, etc.). By analyzing this kind of graph, we can detect suspicious communities or outliers that do not conform to expected patterns such as illegitimate journals and publishers or authors with questionable publishing practices.}
\label{fig-graph-model-overview}
\end{figure}

To overcome these limitations, we argue for a methodology based on multivariate graph analysis. This approach leverages the interconnected nature of the academic publishing ecosystem to provide a more comprehensive understanding of journal legitimacy. By constructing a graph (Figure~\ref{fig-graph-model-overview})that represents the relationships between authors, articles, journals, and publishers, we can employ advanced graph algorithms to detect patterns, communities, centralities, and anomalies within the network.

Next figure (Figure~\ref{fig-paper-centric-detail}) zooms in on a specific subgraph, focusing on two papers and their associated entities: authors and their affiliations, journals, topics and subtopics. It demonstrates the level of detail that can be achieved through this approach. As we can see, the nodes and edges can also include attributes like year of publication, APC fee, journal impact factor, h-index of authors, and more.

\begin{figure}[!htbp]
\centering
\includegraphics[width=0.75\linewidth]{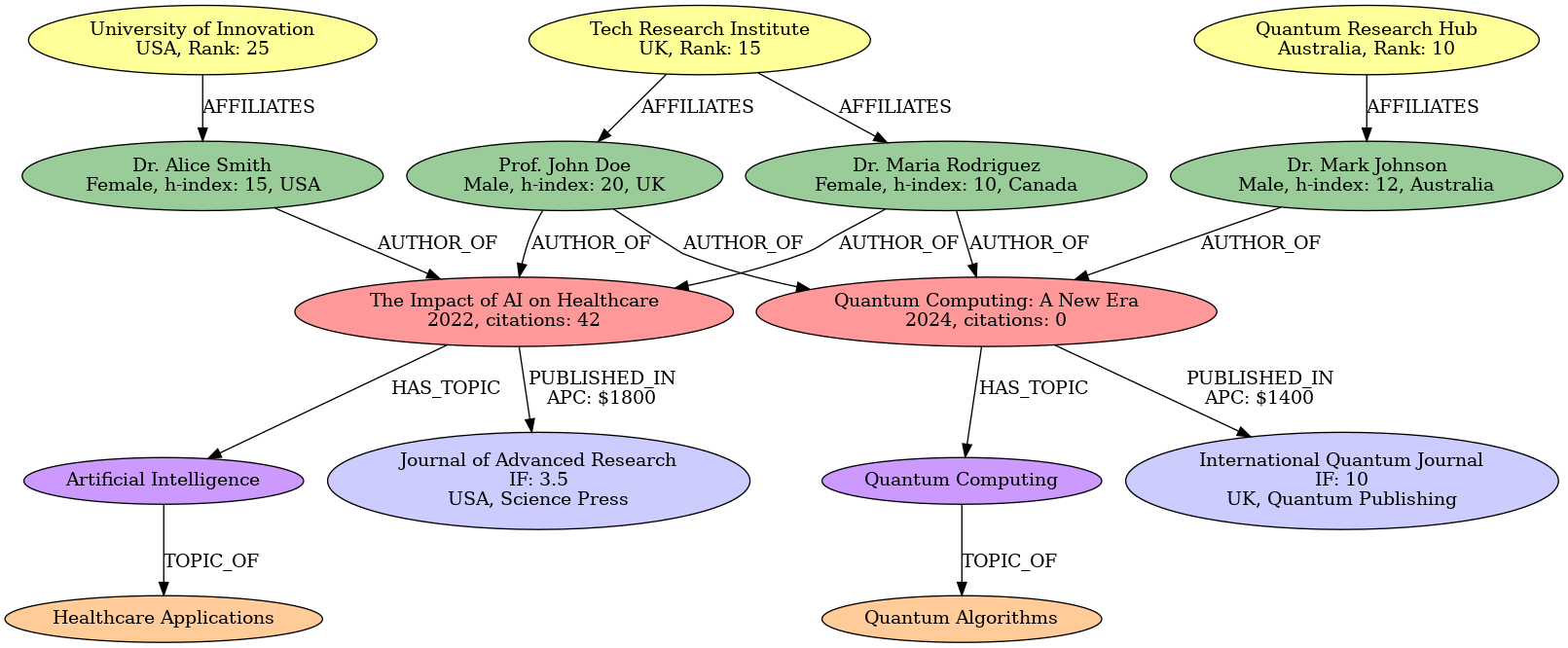}
\caption[]{Zoom-In on a Paper-Centric Detail

This figure demonstrates a detailed subgraph focusing on two papers (red) and their associated entities including attributes.}
\label{fig-paper-centric-detail}
\end{figure}

Potential Use Cases for the Graph-Based Approach:

\begin{itemize}
\item Anomaly Detection: Detecting anomalies in the graph, such as authors with unusual publishing patterns, journals or publishers with circular references, or clusters of journals with suspiciously high citation rates. These anomalies could indicate illegitimate practices and warrant further investigation.
\item Community Detection: Identifying communities of authors, journals, or publishers that are closely connected within the graph. These communities could represent legitimate research groups, reputable journals, or predatory publishers. By analyzing the relationships between these entities, we can gain insights into the structure of the academic publishing ecosystem.
\item Improving Research Integrity: Institutions can use this graph-based model to assess the publication practices of their faculty and students, ensuring that they publish in reputable journals. This method can also help in evaluating the overall research integrity of an institution by identifying patterns of engagement with illegitimate journals. Also identifying (co-)authors who often publish in illegitimate journals can be a valuable insight.
\item Link Analysis: Finding the most influential journals, authors, or publishers in a particular field by analyzing the graph's centrality algorithms or page-rank scores.
\item Enhancing Bibliometric Analysis: Traditional bibliometric indicators like the impact factor can be supplemented with graph-based insights to provide a more nuanced evaluation of journal quality. For example, journals that are closely linked to clusters of known predatory publishers can be flagged as potentially suspicious, even if they have a high impact factor.
\item Supporting Open Access Policies: Policymakers and funding agencies can use the insights derived from graph analysis to support open access policies that promote legitimate publishing practices. By understanding the complex dynamics of the academic publishing network, they can develop more targeted interventions to promote transparency and integrity in scholarly publishing.
\end{itemize}

Crucially, such an approach is now \textit{feasible without proprietary databases}. The open bibliometric index OpenAlex \citep{priemOpenAlexFullyopenIndex2022} provides openly licensed metadata for hundreds of millions of works, and independent validation studies have established it as a viable alternative to Web of Science and Scopus in coverage and quality \citep{thelwallOpenAlexSuitableResearch2025, culbertReferenceCoverageAnalysis2025, alperinAnalysisSuitabilityOpenAlex2024}. This means that any institution can construct and audit the publishing network itself, in line with the principles of reproducible science. This review does not itself validate the graph framework; it identifies the relational gap and motivates the companion empirical artefact that realizes and tests it. In a companion paper we concretize and empirically realize the approach outlined here --- defining a heterogeneous graph model over OpenAlex, implementing it as an open-source library, and demonstrating community detection and a transparent, screening-oriented anomaly analysis on real institutional and thematic corpora [Šamárek \& Martinek, forthcoming]. That work treats structural signals as candidates for human review rather than verdicts, and shows --- through matched controls --- that some intuitive signals do not survive validation.

\section{Conclusion}

After introduction of Open Access publishing and following rise of so called predatory journals, the scientific community has been struggling to define and identify this kind of journals.

While Beall's list initiated the discussion about predatory journals and was a valuable tool in the early days of OA publishing, black/white listing (criteria-based binary classification) is not a suitable approach for identifying illegitimate journals today.

Several approaches have been introduced, but none of them has been universally accepted. The existing methods are either time-consuming and not scalable\citep{grudniewiczPredatoryJournalsNo2019}, or automated but the accuracy, universality, and interpretability of their results are questionable\citep{teixeiradasilvaCanAIDetect2023}.

In this study, we have classified the existing methodologies and compared their strengths and weaknesses. We have identified a gap in the current methods and proposed a new, unexplored approach based on multivariate graph analysis. This method has the potential to provide a complex, more intuitive, transparent, and comprehensive insight into the whole academic publishing network.

To address the journal binary classification dilemma, instead of labeling journals either as legitimate or illegitimate, we propose to use wider range of labels or better quantify the confidence of the journal's legitimacy.

By advancing our ability to surface and deter illegitimate publishing practices, we can help safeguard the integrity of academic publishing and ensure that scholarly communication remains a trusted foundation for knowledge dissemination and advancement.

\section{Future Directions}

The proposed multivariate graph analysis approach offers a promising alternative to traditional methods of identifying illegitimate journals. By leveraging the power of graph algorithms, this methodology can provide a more holistic and intuitive understanding of the academic publishing landscape.

With the core infrastructure and a first realization established in the companion work, future research should focus on scaling the analysis from institutional to national and global corpora, on incorporating the \textit{temporal} dynamics of citation and collaboration (so that emerging coordinated patterns can be detected as they form), and on enriching the structural view with a semantic layer derived from document content. A further priority is rigorous evaluation against curated ground truth --- retraction records, delisted-venue registries, and known paper-mill networks --- together with the matched controls needed to separate genuine integrity signals from confounds such as venue prominence.

\section{Appendix 1}\label{appendix1}

An example of a Manual Analysis based on investigation of journal's processes and journal's description:

\begin{figure}[!htbp]
\centering
\includegraphics[width=0.75\linewidth]{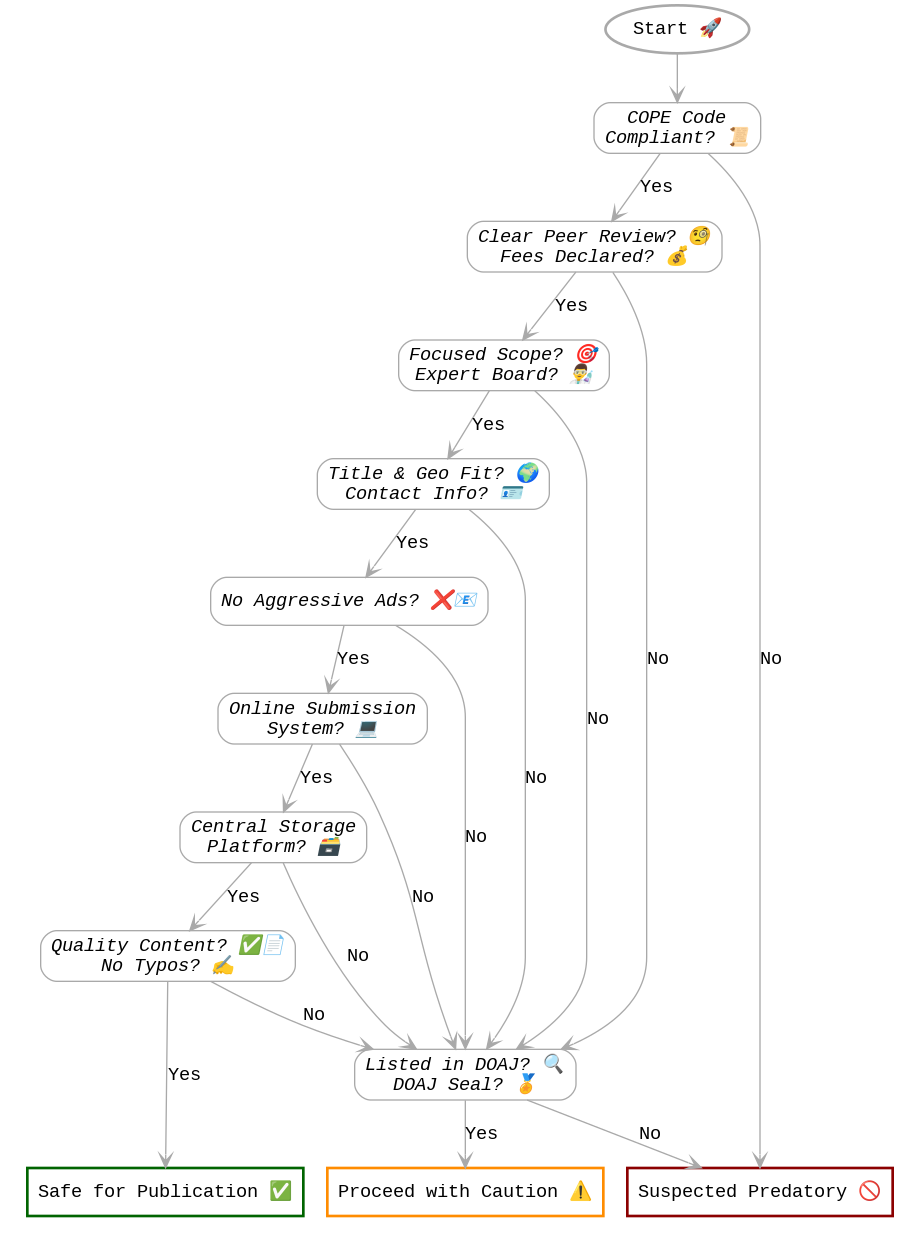}
\caption[]{Manual Analysis Flowchart

This flowchart outlines a step-by-step manual analysis for researchers to determine the trustworthiness of academic journals. It poses critical questions about the journal's practices, from following the Committee on Publication Ethics Code of Conduct \citep{COPE_Code_of_Conduct} to the quality of the published content. The tree leads to three possible outcomes: identifying a journal as safe for publication, suggesting caution, or suspecting predatory practices.

Inspired by \citep{richtigProblemsChallengesPredatory2018}.}
\label{manual-analysis-flowchart}
\end{figure}

\bibliography{main}
\end{document}